# Three-dimensional Electro-convective Vortices in Cross-flow

Yifei Guan[1], James Riley[1], and Igor Novosselov [1,2,§]

[1] *Department of Mechanical Engineering, University of Washington, Seattle, U.S.A. 98195*

[2] *Institute for Nano-Engineered Systems, University of Washington, Seattle, U.S.A. 98195*

Aug 2019

The study focuses on the 3D electro-hydrodynamic (EHD) instability for flow between to parallel electrodes with unipolar charge injection with and without cross-flow. Lattice Boltzmann Method (LBM) with two-relaxation time (TRT) model is used to study flow pattern. In the absence of cross-flow, the base-state solution is hydrostatic, and the electric field is one-dimensional. With strong charge injection and high electrical Rayleigh number, the system exhibits electro-convective vortices. Disturbed by different perturbation patterns, such as rolling pattern, square pattern, and hexagon pattern, the flow patterns develop according to the most unstable modes. The growth rate and the unstable modes are examined using dynamic mode decomposition (DMD) of the transient numerical solutions. The interactions between the applied Couette and Poiseuille cross-flows and electroconvective vortices lead to the flow patterns change. When the cross-flow velocity is greater than a threshold value, the spanwise structures are suppressed; however, the cross-flow does not affect the streamwise patterns. The dynamics of the transition is analyzed by DMD. Hysteresis in the 3D to 2D transition is characterized by the non-dimensional parameter *Y*, a ratio of the coulombic force to viscous term in the momentum equation. The change from 3D to 2D structures enhances the convection marked by a significant increase in the electric Nusselt number.

## I. INTRODUCTION

Both 2D and 3D vortex structures are ubiquitous in fluid systems. In considering convection, various flow patterns result from a body force acting on the fluid, e.g., Rayleigh-Benard convection (RBC) [1-7], Marangoni effects [8-12], magneto-convection [13-21], and magnetohydrodynamics convection [22-29]. Electro-convection (EC) phenomenon has been first reported by G. I. Taylor in 1966 describing cellular convection in the liquid droplet [30]. Since then, EC has been observed in other systems with the interaction of electric force with fluids. In nonequilibrium electro-hydrodynamic (EHD) systems [7, 30-50], poorly conductive leaky dielectric fluid acquire unipolar charge injection at in the surface interface in response to the electric field. In charge-neutral electro-kinetic (EK) systems, electro-convection is triggered by the electro-osmotic slip of electrolyte in the electric double layer at membrane surfaces [51-62]. The transition from 3D to 2D structures under the influence of shear stress has been observed in atmospheric cloud streets in the planetary boundary layers [63-65], and in laboratory RBC studies [66-70]. In EC systems [7, 32, 33, 35, 38-50, 56, 59, 62, 71] the forcing term is different from that in the other systems, both 3D and 2D patterns can exist; these are determined by the balance of the forces acting on the fluid.

The EC stability problem was first analyzed by a reduced non-linear hydraulic model [72, 73] and by a linear stability analysis without the charge diffusion term [74, 75]. Atten & Moreau [76] showed that, in the weak-injection limit, C<<1, the flow stability is determined by the parameter $T_c C^2$, where *C* is the charge injection level and $T_c$ is the linear stability threshold for the electrical Rayleigh number $T$, the ratio of electrical force to the viscous force (Eq. (9)). In the space-charge-limited (SCL) injection ($C \to \infty$), the flow stability depends on $T_c$ alone [77-79]. The effect of charge diffusion was investigated by Zhang et al. by employing linear stability analysis on the EC problem with a Poiseuille flow [42], and by non-linear analysis using a multiscale method [47]. The authors found that the charge diffusion has a non-negligible effect on $T_c$, and the transient behavior depends on the Reynolds number ($Re$) [42, 47]. Li et al. performed linear analysis to study convective instabilities in EHD-Poiseuille flow and found that the ratio of the Coulomb force to the viscous force has an impact on the transition of transverse rolls from convective to absolute instability [80].

§ ivn@uw.edu

Though the 3D EC stability problem in the presence of shear is complicated, it is somewhat analogous to Rayleigh-Bernard convection (RBC) [66-70, 81-84]. For example, in cross-flow, the suppression of the transverse cells, and the evolution of the longitudinal cells has been reported; in the meteorological application, the addition of cross-flow leads to the formation of cloud streets [63]. Mohamad et al. used a non-dimensional group $Gr/Re^2$, the ratio of buoyancy to the inertia force, to parametrize the effect of applied shear, where $Gr$ is the Grashof number [70]. For $Gr/Re^2 > 10$, the impact of the cross-flow is insignificant, while for $Gr/Re^2 < 0.1$, the effect of the buoyancy can be neglected. Reduced nonlinear models such as Ginzburg-Landau equations are used to study the transitional behavior of RBC cells [66-69].

Numerical simulation can be used to shed insight into the behavior of EC vortices. Early work has shown that infinite-difference modeling, strong numerical diffusivity can lead to an incorrect prediction of stability criteria [33]. Other numerical approaches have been developed, including the particle-in-cell method [85], the finite-volume method with flux-corrected transport [86] or the total variation diminishing scheme [38, 40, 44-46], and the method of characteristics [35]. Luo et al. showed that a unified Lattice Boltzmann model (LBM) predicts the linear and finite-amplitude stability criteria of the subcritical bifurcation in the EC flow for both 2D and 3D flow scenarios [7, 48-50, 71]. This unified LBM transforms the elliptic Poisson equation into a parabolic reaction-diffusion equation and introduces artificial coefficients to control the evolution of the electrical potential. A segregated solver was proposed that combines a two-relaxation time (TRT) LBM modeling of the fluid and charge transport, and a Fast Fourier Transform (FFT) Poisson solver for the electrical field [87]. Related to vortex structure transition with the addition of cross-flow, 2D finite-volume simulations of Poiseuille flow have demonstrated that the value of $T_c$ is a function of $Re$ and the ion mobility parameter, $M$ [43]. More recently, 2D numerical simulations have been used to parameterize the flow transition of EC vortex pairs to the base cross-flow [88].

Among other methods, dynamic mode decomposition (DMD) has been used to analyze the behavior of a complex flow system. DMD is a data-driven analysis performed on experimental measurements or numerical solutions, shedding insight into the spatiotemporal dynamics of complex systems [89]. Schmid and colleagues first applied DMD to the stability analysis of fluid flow [90, 91]. The eigenmodes from DMD are equivalent to global modes if the linearized equations are used in numerical simulations. DMD has also been used to identify bifurcation points in complex systems such as flow in a lid-driven cavity of high Reynolds number [92], to reconstruct compressed high-dimensional data of a fluid system [93], and to extract coherent spatiotemporal structures in fluid flows for prediction and control [94, 95]. DMD for EC flow instability analysis, however, has not been reported.

The current work studies the transitional behavior of EC vortices in the presence of cross-flow using a segregated TRT LBM solver. Couette or Poiseuille cross-flow is added to EC convection before and after the vortices are established. DMD analysis of the numerical solution sheds insight into the formation and pattern transitions of 2D and 3D coherent fluid structures. The effects of cross-flow are parameterized by a non-dimensional number, *Y*, a ratio of electrical to viscous forces.

## II. GOVERNING EQUATIONS AND DIMENSIONAL ANALYSIS

The governing equations for the system are the Navier-Stokes equations (NSE) with an electrical forcing term $\mathbf{F_e} = -\rho_c \nabla \varphi$ added to the momentum equation, the charge transport equation, and the Poisson equation for electrical potential.

$$\nabla \cdot \mathbf{u} = 0, \quad (1)$$

$$\rho \frac{D\mathbf{u}}{Dt} = -\nabla P + \mu \nabla^2 \mathbf{u} - \rho_c \nabla \varphi, \quad (2)$$

$$\frac{\partial \rho_c}{\partial t} + \nabla \cdot \left[ (\mathbf{u} - \mu_b \nabla \varphi) \rho_c - D_c \nabla \rho_c \right] = 0, \quad (3)$$

$$\nabla^2 \varphi = -\frac{\rho_c}{\varepsilon}, \quad (4)$$

where $\rho$ and $\mu$ are the density and the dynamic viscosity of the working fluid, $\mathbf{u} = (u_x, u_y, u_z)$ is the velocity vector field, $P$ is the static pressure, $\rho_c$ is the charge density, $\mu_b$ is the ion mobility, $D_c$ is the ion diffusivity, $\varepsilon$ is the electrical permittivity, and $\varphi$ is the electrical potential. The electrical force is a source term in the momentum equation (Eq. (2)) [42, 96-98]. The variables to be solved are velocity field $\mathbf{u}$, pressure $P$, charge density $\rho_c$, and electrical potential $\varphi$. The flow is assumed to be periodic in the x- and y-directions, and wall-bounded z-direction. Cross-flow is applied in the x-direction.

In the absence of cross-flow, the system can be non-dimensionalized with the characteristics of the electric field [42]: $H$ is the distance between the electrodes (two plates infinite in x and y), $\rho_0$ is the injected charge density at the anode, and $\Delta \varphi_0$ is the voltage difference applied to the electrodes. The time $t$ is non-dimensionalized by $H^2 / (\mu_b \Delta \varphi_0)$, the velocity $\mathbf{u}$ -- by the drift velocity of the ions $u_{drift} = \mu_b \Delta \varphi_0 / H$, the pressure $P$ -- by $\rho_0 (\mu_b \Delta \varphi_0)^2 / H^2$, and the charge density $\rho_c$ by $\rho_0$. Therefore, the non-dimensionalization of the governing equations (Eq. (1)-(4)) gives:

$$\nabla^* \cdot \mathbf{u}^* = 0, \quad (5)$$

$$\frac{D^* \mathbf{u}^*}{D^* t^*} = -\nabla^* P^* + \frac{M^2}{T} \nabla^{*2} \mathbf{u}^* - CM^2 \rho_c^* \nabla^* \varphi^*, \quad (6)$$

$$\frac{\partial^* \rho_c^*}{\partial^* t^*} + \nabla^* \cdot \left[ (\mathbf{u}^* - \nabla^* \varphi^*) \rho_c^* - \frac{1}{Fe} \nabla^* \rho_c^* \right] = 0, \quad (7)$$

$$\nabla^{*2} \varphi^* = -C \rho_c^*, \quad (8)$$

where the asterisk denotes non-dimensional variables. These non-dimensional equations yield four dimensionless parameters describing the system's state [7, 38-50].

$$M = \frac{(\varepsilon / \rho)^{1/2}}{\mu_b}, \quad T = \frac{\varepsilon \Delta \varphi_0}{\mu \mu_b}, \quad C = \frac{\rho_0 H^2}{\varepsilon \Delta \varphi_0}, \quad Fe = \frac{\mu_b \Delta \varphi_0}{D_e}. \quad (9)$$

The physical interpretations of these four non-dimensional parameters are as follows: $M$ is the mobility ratio between hydrodynamic mobility and the ionic mobility; $T$ is the electrical

Rayleigh number, a ratio between the electrical force and the viscous force; $C$ is the strength of injection [42, 47]; and $Fe$ is the reciprocal charge diffusivity coefficient [42, 47, 80].

With the addition of a cross-flow, the velocity term in the non-dimensional momentum equation is modified to account for the external flow, $\mathbf{u}_{ext}$, which is different from the previous formulations where the drift charge velocity was used [88]. Here we use the velocity of the upper wall in Couette flow or the centerline velocity for Poiseuille flow as $\mathbf{u}_{ext}$ [88]. The non-dimensional momentum and charge transport equations become:

$$\frac{D^*\mathbf{u}^*}{D^*t^*} = -\nabla^* P^* + \frac{1}{\text{Re}}\nabla^{*2}\mathbf{u}^* - X\rho_c^*\nabla^*\varphi^*, \tag{10}$$

$$\frac{\partial^* \rho_c^*}{\partial^* t^*} + \nabla^* \bullet \left[\left(\frac{|\mathbf{u}_{ext}|}{u_{drift}}\mathbf{u}^* - \nabla^*\varphi^*\right)\rho_c^* - \frac{1}{Fe}\nabla^*\rho_c^*\right] = 0, \tag{11}$$

where the Reynolds number is $Re = \frac{\rho |\mathbf{u}_{ext}| H}{\mu}$, and $X = \frac{\rho_0 \Delta\varphi_0}{\rho |\mathbf{u}_{ext}|^2}$ is the ratio of the electrical force to the inertial force [97]. Since $Re$ is essentially the ratio of inertia to viscous force, and $X$ is the ratio of electrical force to inertia, the product of these (denoted as $Y$) is the ratio of the electrical force to the viscous force:

$$Y = X \times \text{Re} = \frac{\rho_0 \Delta\varphi_0 H}{\mu |\mathbf{u}_{ext}|} = \frac{\rho_0 \Delta\varphi_0}{|\boldsymbol{\tau}|}, \tag{12}$$

where $\boldsymbol{\tau}$ is the shear stress $\boldsymbol{\tau} = \mu \frac{\mathbf{u}_{ext}}{H}$. In Couette flow $\boldsymbol{\tau} = constant$ ($\mathbf{u}_{ext} = u_{wall}\mathbf{e}_x$); while in Poiseuille flow, the average value in the channel flow is used hereafter ($\mathbf{u}_{ext} = u_{center}\mathbf{e}_x$ and $H$ replaced by half-height $H/2$), where $\mathbf{e}_x$ is a unit vector in the x-direction. Detailed analysis of Y for 2D EC flow can be found in the previous report [88].

### III. SYSTEM LINEARIZATION AND INITIALIZATION

In the flow stability analysis problems, the initial linear growth region can be described by the linearized governing equations. The dimensional variables can be written as a summation of the base state (denoted with an overbar) and perturbation (denoted with prime), i.e., $u = \bar{u} + u'$, $P = \bar{P} + P'$, $\varphi = \bar{\varphi} + \varphi'$, and $\rho_c = \bar{\rho}_c + \rho_c'$ [42]. The base state variables are only functions of $z$. Substituting these expressions into Eq. ((1)-(4)), subtracting the governing equations for the base states, and truncating the second-order perturbation terms yields the linear system:

$$\nabla \bullet \mathbf{u}' = 0, \tag{13}$$

$$\rho \frac{\partial \mathbf{u}'}{\partial t} + (\mathbf{u}' \bullet \nabla)\bar{\mathbf{u}} + (\bar{\mathbf{u}} \bullet \nabla)\mathbf{u}' = -\nabla P' + \mu \nabla^2 \mathbf{u}' - \left(\rho_c' \nabla \bar{\varphi} + \bar{\rho}_c \nabla \varphi'\right), \tag{14}$$

$$\frac{\partial \rho_c}{\partial t} + \nabla \bullet \left[\left(\bar{\mathbf{u}} - \nabla\bar{\varphi}\right)\rho_c' + \left(\mathbf{u}' - \nabla\varphi'\right)\bar{\rho}_c - D_c \nabla \rho_c'\right] = 0, \tag{15}$$

$$\nabla^2 \varphi' = -\frac{\rho_c'}{\varepsilon}, \tag{16}$$

which can be written symbolically as

$$\frac{d\gamma}{dt} = \mathbf{L}\gamma, \tag{17}$$

where $\gamma$ is the vector of unknowns, and $\mathbf{L}$ is the linear differential operator.

For periodic boundary conditions in the x- and y-directions, the normal modes take the form

$$\gamma = W(z)f(x,y)e^{\omega t}, \tag{18}$$

where $\gamma$ represents any flow variable ($\mathbf{u}', p', \rho_c', \varphi'$); $\omega$ is the eigenvalue of the spatial-differential matrix $\mathbf{L}$, and $W(z)f(x,y)$ is the corresponding eigenfunction. The choice of the normal modes depends on the initial perturbation (initial conditions).

To initialize the system, the hydrostatic base state is obtained, as shown in FIG. 1. Without initial perturbation, the system is hydrostatic, and the electrical properties are one-dimensional in the z-direction.

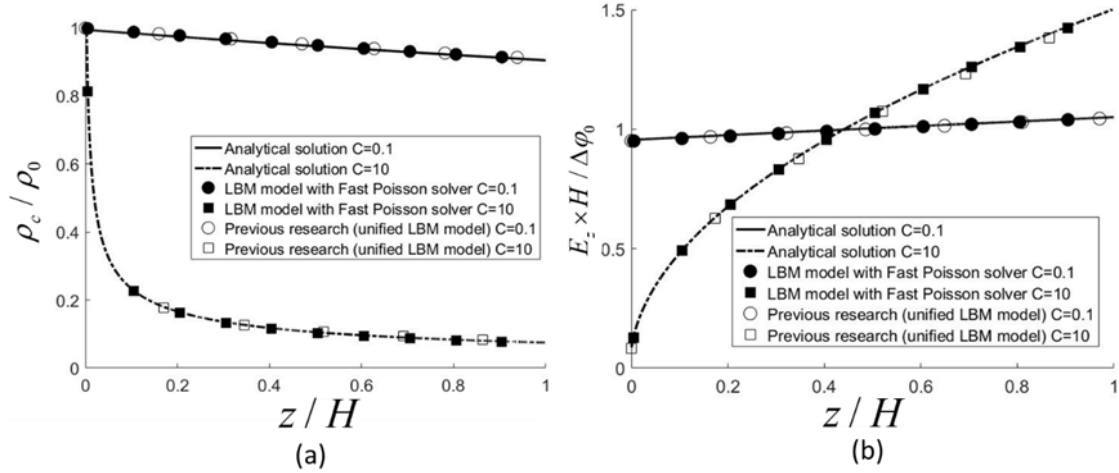

**FIG. 1.** Hydrostatic solution comparison of the TRT LBM and Fast Poisson solver [87, 88], unified SRT LBM [48], and the analytical solution [85, 98] for $C=0.1$ and $C=10$, $Fe=4000$. (a) Electric field and (b) charge density;

To obtain various equilibrium solutions, for example, as shown in FIG. 2, different initialization (initial perturbation) schemes are applied to the hydrostatic base state. The initial perturbation used in the simulations has a form similar to the eigenfunction of the normal mode $W(z)f(x,y)$. To satisfy the continuity condition (Eq. (13)), the initial velocity field is described as

$$u_z = W(z)f(x,y),\ u_x = \frac{1}{a^2}\frac{\partial^2 u_z}{\partial x \partial z}, u_y = \frac{1}{a^2}\frac{\partial^2 u_z}{\partial y \partial z}, \tag{19}$$

where $a$ depends on the wavelengths in x- and y-directions and satisfies

$$\left(\frac{\partial^2}{\partial x^2} + \frac{\partial^2}{\partial y^2}\right)f(x,y) = -a^2 f(x,y). \tag{20}$$

The initial perturbation for a rolling pattern (2D) is taken to be:

$$u_x = 0, \tag{21}$$

$$u_y = -\frac{dW(z)}{dz}\frac{1}{a^2}\frac{2\pi}{L_y}\sin(2\pi y/L_y), \tag{22}$$

$$u_z = W(z)\cos(2\pi x/L_y), \tag{23}$$

$W(z)$ is chosen to satisfy the no-slip boundary conditions at the walls. $L_y$ is the wavelength in the y-direction (spanwis. The initial perturbation for a square pattern (3D) is taken to be:

$$u_x = -\frac{dW(z)}{dz}\frac{1}{a^2}\frac{2\pi}{L_x}\sin(2\pi x/L_x), \tag{24}$$

$$u_y = -\frac{dW(z)}{dz}\frac{1}{a^2}\frac{2\pi}{L_y}\sin(2\pi y/L_y), \tag{25}$$

$$u_z = W(z)\left[\cos(2\pi x/L_x) + \cos(2\pi y/L_y)\right]. \tag{26}$$

For the square patterns $L_x = L_y$, from Eq. (20):

$$a = 2\pi/L_y. \tag{27}$$

The initial perturbation for the hexagon pattern (3D) is taken to be:

$$u_x = -\frac{dW(z)}{dz}\frac{4\pi}{3\sqrt{3}Lc^2}\sin\left(\frac{2\pi x}{\sqrt{3}L}\right)\cos\left(\frac{2\pi y}{3L}\right), \tag{28}$$

$$u_y = -\frac{dW(z)}{dz}\frac{4\pi}{9Lc^2}\left[\cos\left(\frac{2\pi x}{\sqrt{3}L}\right) + 2\cos\left(\frac{2\pi y}{3L}\right)\right]\sin\left(\frac{2\pi y}{3L}\right), \tag{29}$$

$$u_z = \frac{1}{3}W(z)\left[2\cos\left(\frac{2\pi x}{\sqrt{3}L}\right)\cos\left(\frac{2\pi y}{3L}\right) + \cos\left(\frac{4\pi y}{3L}\right)\right], \tag{30}$$

where $L$ is the side of the hexagon and parameter $c = \frac{4\pi}{3L}$ to satisfy Eq. (20).

To satisfy the wall-bounded no-slip boundary condition in the z-direction, we use

$$W(z) = \left[\cos(2\pi z/H) - 1\right]\times\varepsilon, \tag{31}$$

where $\varepsilon = 10^{-3}$ is the perturbation magnitude, taken to be the same as in previous 2D analyses [87, 88].

## IV. DYNAMIC MODE DECOMPOSITION

To study the coherent structures leading to flow instability, we perform DMD on the numerical data for $u_z$. DMD reconstructs the complex flow system using the linear growth approximation between snapshots of numerical solutions [89]; DMD examines the coherent flow structures and can be used as a tool for flow field predictions and stability analysis. A continuous linear dynamical system (Eq. (17)) can be described by an analogous time-discretized system at intervals $\Delta t$:

$$\gamma_{k+1} = \mathbf{A}\gamma_k, \tag{32}$$

where

$$\mathbf{A} = \exp(\Delta t \mathbf{L}), \tag{33}$$

and $\gamma_k$ is any flow variable ($\mathbf{u}', p', \rho_c', \varphi'$) at a time step $k$. The operator $\mathbf{L}$ is a spatial differential matrix of the continuous-time dynamical system as in Eq. (17). The solution to the discrete-time system can be expressed in terms of eigenvalues $\lambda_j$ and corresponding eigenvectors $\xi_j$ of the discrete-time mapping matrix $\mathbf{A}$:

$$\gamma_k = \sum_{j=1}^{r} \xi_j \lambda_j^k b_j = \xi \mathbf{\Lambda}^k \mathbf{b}, \tag{34}$$

where $\mathbf{b}$ contains the coefficients of the perturbation (initial conditions) $\gamma_1$ in the eigenvector basis, such that $\gamma_1 = \xi \mathbf{b}$, $r$ is the rank of the reduced eigenmodes, $\xi$ is the matrix whose columns are the eigenvectors $\xi_j$, and $\mathbf{\Lambda}^k$ is a diagonal matrix whose entries are the eigenvalues $\lambda_j$ raised to the power of $k$.

The results obtained by the DMD algorithm based on the data collected from the numerical simulations can be compared to the values calculated by linear stability analysis in the linear growth region.

With the low-rank approximation of both the eigenvalues and the eigenvectors, the projected future solution can be constructed.

$$\boldsymbol{\gamma}(t) \approx \sum_{j=1}^{r} \xi_j \exp(\omega_j t) b_j = \xi \exp(\mathbf{\Omega} t) \mathbf{b}, \tag{35}$$

where $\omega_j = \ln(\lambda_j)/\Delta t$ and $\mathbf{b} = \xi^\dagger \boldsymbol{\gamma}_1$, $\xi$ is the matrix whose columns are the DMD eigenvectors $\xi_j$, superscript $\dagger$ is the Moore-Penrose pseudoinverse and $\mathbf{\Omega}$ is a diagonal matrix whose entries are the eigenvalues $\omega_j$. The details of the implementation of DMD is included in the supplementary materials.

## V. RESULTS AND DISCUSSION

The TRT LBM approach is used to solve the transport equations for fluid flow and charge density, coupled to a fast Poisson solver for electric potential [87, 88]. The solver is extended to 3D for the differential equations (Eq. (1)-(4)), the simulations are performed using the initial perturbations (i.e., initial conditions) as per Eq. (21)-(23) for the rolling pattern, Eq. (24)-(26) for the square pattern, and Eq. (28)-(30) for the hexagon pattern and mixed patterns. The equilibrium state was obtained when the flow became steady. The numerical code implementation is in SI units, and the physical constants are determined by the non-dimensional parameters. The numerical method is implemented in C++ using CUDA GPU computing. FFT and IFFT operations use the cuFFT library [99-101]. All variables are computed with double precision to reduce truncation errors. The numerical method was shown to be 2nd order accurate in space [87]. Error analysis is provided in supplementary materials.

### *1. Electro-convection vortices and transition to rolls: General patterns*

The equilibrium patterns of EC flow fields without cross-flow were obtained using the initial perturbations described in section III. The resulting patterns depend on the non-dimensional parameter $T$ and the domain size; the latter determines the wavelengths of the vortices. FIG. 2(a-e) shows the equilibrium states of $u_z$ at $z = H/2$. The values $C = 10$, $M = 10$, and $Fe = 3500$ were held constant for each condition. The simulations were carried at $T = 170$ for $u_z$ plotted in FIG. 2(a, c-e), and to $T = 833$ for $u_z$ in FIG. 2(b). The domain sizes and initial perturbation for the simulations plotted in FIG. 2(a) and (b) are the same (Eq. (24)-(26)), and therefore $u_z$ in FIG. 2(b) is the harmonic of $u_z$ in FIG. 2(a) develops at a higher value of $T$. For cases given in FIG. 2(c-e), different domain sizes with hexagon initial perturbation (Eq. (28)-(30)) were used. When sufficiently strong Couette type cross-flow in the x-direction was applied to the 3D structures, the transition to 2D streamwise rolling patterns occurs for all initial perturbations scenarios (FIG. 2(f-j)).

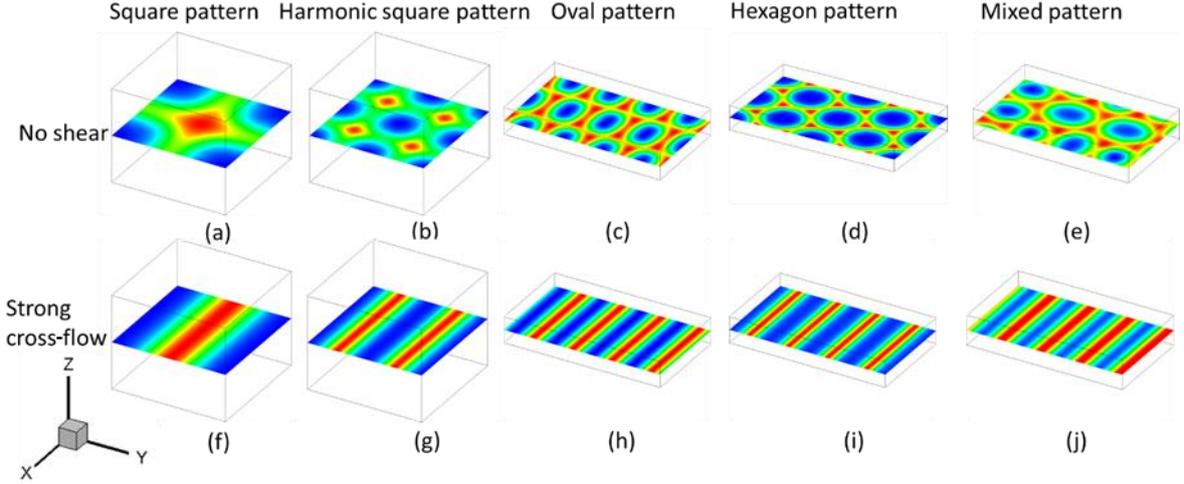

**FIG. 2.** Contours of $u_z$ at $z = H/2$ at equilibrium states (a-e) without cross-flow, and (f-j) with cross-flow sufficient for pattern transition. For different electrical Rayleigh numbers, domain sizes, and initial perturbations (initial conditions), square patterns, oval patterns, hexagon patterns, and mixed patterns are established. Strong cross-flow in the x-direction is applied to the equilibrium states (a-e), resulting in the 3D transition to 2D streamwise vortices.

To study the mechanism for the transition of 3D vortices to streamwise vortices, we consider the simplest scenario, i.e., the case where the equilibrium state is a single period square pattern, see FIG. 2(a). Further generalization of transition for other patterns can be a subject of future work. The physical domain used in the simulation is given by $L_x = L_y = 1.22 m$ and $H = 1 m$; this limits the wavenumber to $k_x = k_y = 2\pi / L_x \approx 5.15 (1/m)$ [42, 49]. The electrical Nusselt number, $Ne = I / I_0$, serves as a flow stability criteria, where $I$ is the cathode current for a given solution and $I_0$ is the cathode current for the base state solution without EC vortices [38, 49]; thus, $Ne > 1$ when EC vortices exist. Note that the use of current as the metric for EC convection has been used in the studies of related overlimiting current in electro-kinetic systems [48]. The transition to EC chaotic flow [59] at higher values of the forcing term is not considered in this paper. Since, for the EC problem with the cross-flow, the stability largely depends on $Y$ [88], in this analysis $Y$ is varied, while other non-dimensional parameters are held constant at $T = 170$, $C = 10$, $M = 10$, and $Fe = 3500$.

### 2. *Square and rolling patterns Perturbation of the hydrostatic base state*

An initial perturbation was applied to the hydrostatic base state after the one-dimensional electrical property profiles were established as shown in FIG. 1. FIG. 3 shows that for $T > T_c$, the 2D perturbation (Eq. (21)-(23)) leads to the development of a rolling pattern with flow only in the y and z directions, while the 3D square perturbation (Eq. (24)-(26)) leads to a square pattern with velocities in all three directions. FIG. 4 shows the evolution of the maximum $u_z$ for the first $t^*$=0~5 after the perturbation is applied (non-dimensional time scale is obtained via normalization by $H^2/(\mu_b \Delta \varphi_0)$, as described in Section II). Both the rolling pattern and square pattern have the same non-dimensional linear growth rate (~0.896 as $\omega$ in Eq. (18)), which agrees with the previously reported linear stability analysis [42] and the unified SRT LBM numerical model [49] (the case with $T=170$, $C=10$, $M=10$, and $Fe=4000$ is included in supplementary materials for validation). After about $t^*=5$ from the initial perturbation, the growth rate curves for square and rolling vortices patterns diverge. Although the maximum $u_z^*$ is greater for the square pattern, the charge transport (based on $Ne=I/I_0$) for rolling patterns is greater, as shown later in FIG. 17 and FIG. 18.

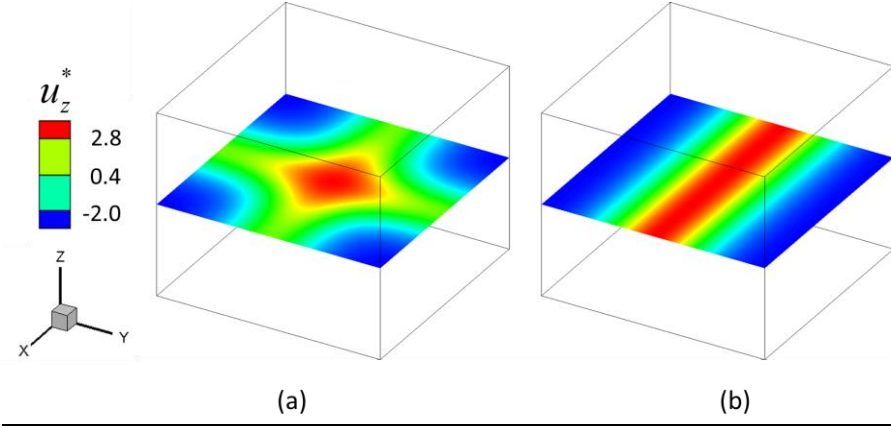

(a)          (b)

**FIG. 3.** Contours of $u_z^*$ at z=H/2 for initial perturbation of (a) a square pattern and (b) a rolling pattern with $T=170$, $C=10$, $M=10$, and $Fe=3500$.

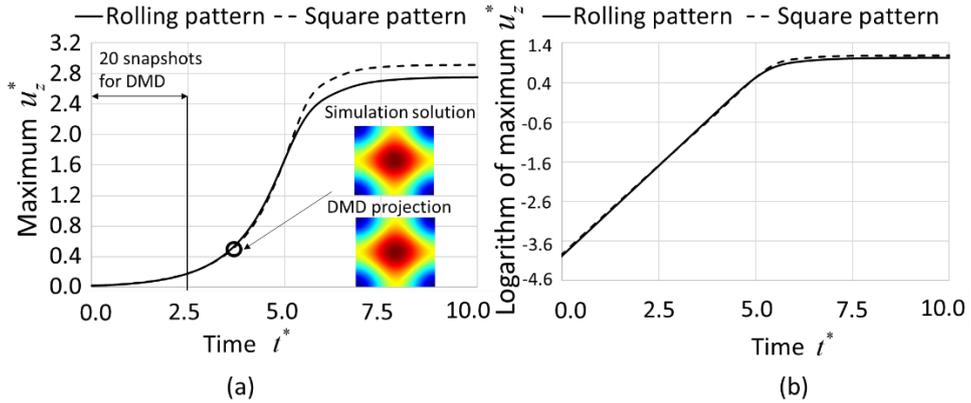

(a)          (b)

**FIG. 4.** Time evolution of maximum $u_z^*$ for the rolling pattern and square pattern in (a) a linear scale and (b) a logarithmic scale. Both patterns have similar growth rates (~0.896) in the linear growth region ($t^*$=0~5). The DMD algorithm-based solutions in the interval $t^*$=0-2.5 project the state at $t^*$=3.75 using Eq. (35).

DMD is performed based on the numerical data of the square pattern perturbation case from $t^*$=0-2.5 at intervals of $\Delta t^* = 0.125$. FIG. 5 shows the eigenvalues $\lambda$ of the discrete-time mapping matrix $\mathbf{A}$ as in Eq. (33) and the logarithmic mapping of the eigenvalues $\omega$ of the matrix $\mathbf{L}$ as in Eq. (17). The eigenvalues $\lambda$ are shown in relationship to the unit circle (dashed line); most of the values are inside the circle and therefore represent stable dynamic modes. Three unstable modes (solid dots) with positive growth rates are found. The unstable modes with $\lambda_i = 0$ suggest that these modes do not oscillate.

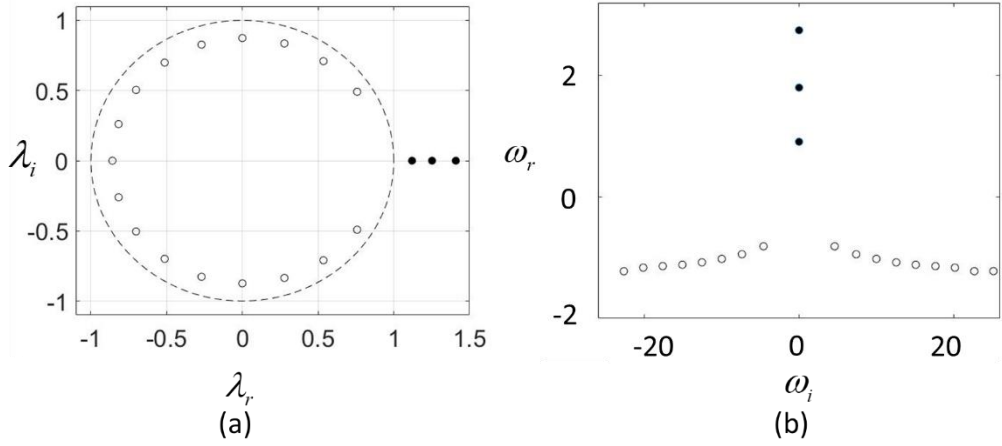

**FIG. 5.** (a) Eigenvalues of the discrete-time mapping matrix $\mathbf{A}$ and (b) logarithmic mapping of eigenvalues of $\mathbf{L}$. The eigenvalues outside the unit circle, whose logarithmic value has a real component $\omega_r$ greater than 0, represent the unstable dynamic modes. The logarithmic mapping of the eigenvalue $\mathbf{L}$ indicates the growth rate of each dynamic mode.

The three unstable modes with positive growth rates dominate the flow pattern. FIG. 6 shows the eigenvectors of these three modes at $t^*$=2.5, plotted on the plane $z = H/2$. The plots show square patterns with different wavelengths and the phase shifts. Although the mode $\omega_r = 0.908$ has a lower growth rate, it contains >99% of the energy of the perturbation (its initial amplitude of b=1.826 is much greater than that of the others). The overall growth rate (~0.896) from FIG. 4 is very close to the growth rate of this dynamic mode shown in FIG. 6(c). The comparison of the dynamic modes FIG. 6(a-c) and simulation solution FIG. 6(d) verifies that the dynamic mode $\omega_r = 0.908$ dominates the flow system. The growth rates of rolling and square patterns are similar to each other, which was also observed in the previous report [49].

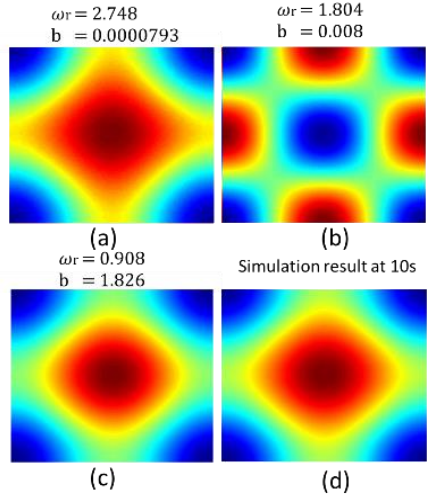

|   |   |
|---|---|
| ωr = 2.748  b = 0.0000793  (a) | ωr = 1.804  b = 0.008  (b) |
| ωr = 0.908  b = 1.826  (c) | Simulation result at 10s  (d) |

**FIG. 6.** Unstable dynamic modes visualized by $u_z$ at $z = H/2$. The dynamic mode $\omega_r = 0.908$ (c) perturbation has the greatest of projection (b=1.826) on this eigenmode; therefore, it contains most of the energy of the system. The growth rate of this mode $\omega_r = 0.908$ is close to the general growth rate of the entire system (~0.896) observed in FIG. 4.

### 3. *Perturbation of the hydrostatic base state solution including a cross-flow*

In this second case, we investigate the EC problem with an initial perturbation applied to the hydrostatic base state after a cross-flow field in the x-direction is developed. Cases for two cross-flows are studied. Only the square initial perturbation (Eq. (24) - Eq. (26)) is considered. Couette flow is obtained by applying the speed $u_{wall}$ to the upper wall, while holding the bottom wall fixed. Poiseuille flow driven by a body force representing the pressure drop - $F_p$, so the center plane velocity is $u_{center} = \frac{1}{2\mu}\left(\frac{H}{2}\right)^2 F_p$. The values of $u_{wall}$ and $u_{center}$ are both non-dimensionalized by $u_{drift}$ such that $u^*_{wall} = u_{wall}/u_{drift}$ and $u^*_{center} = u_{center}/u_{drift}$.

FIG. 7 shows the evolution of maximum $u^*_z$ and *Ne* for both cases. For $t^* < 5$ after the initial perturbation (Eq. (24) - Eq. (26)), the growth is linear, and the growth rate of ~0.896 is the same for all solutions. The growth rate is the same with and without cross-flow because the cross-flow does not affect the streamwise vortices and, as is shown in FIG. 4, the streamwise vortices grow at the same rate as the 3D square patterns. Previously reported linear stability analysis [80] predicts that the cross-flow does not affect the growth rate of longitudinal rolls, but that it decreases the growth rate of traverse rolls. The effect of cross-flow on the growth rate of the traverse rolls is also observed from the simulations, shown in the supplementary material. When the square pattern initial perturbation is applied, both longitudinal and traverse rolls coexist. Since the convolution between these orthogonal rolls decreases the growth rate of each pattern (square and rolling patterns have the same growth rate, as shown in **FIG. 4**), the weakening effect of the cross-flow on the traverse rolls may be compensated by the longitudinal rolls.

At time $t^* \sim 5$, the growth rate curves diverge to reach different equilibrium states for different cross-flow scenarios and perturbation schemes. With weak cross-flow ($u^*_{wall}$=1.6 for Couette flow, $u^*_{center}$=1.68 for Poiseuille flow), the final solutions exhibit oblique 3D vortex structures; both transverse and regular longitudinal rolls coexist. The maximum values of $u^*_z$ of these oblique 3D vortices are greater than for cases with rolling patterns. For strong cross-flow ($u^*_{wall}$=4 for Couette and $u^*_{center}$=3.84 for Poiseuille flow), the systems develop directly into a longitudinal rolling pattern regardless of the initial perturbation; i.e., transverse rolls do not exist even when they are included in the initial perturbation contained in the square pattern. The maximum $u_z$ of the streamwise vortices in the cross-flow case is the same as for the 2D rolling vortices without cross-flow, as shown in FIG. 4; in other words, streamwise vortices are superimposed onto the base state cross-flow solution. For the final steady-state (oblique 3D or 2D rolling vortices), the solutions with and without cross-flow bifurcate at $u^*_{wall}$=2.2 for Couette and $u^*_{center}$=2.8 for Poiseuille flow at about $t^*$=5. Before reaching an equilibrium state, the cases with the moderate cross-flow exhibit an intermediate state where the maximum $u^*_z$ can be greater than the final longitudinal rolling pattern case ($u^*_z$=2.75) or even the square pattern case with both transverse and longitudinal rolls ($u^*_z$=2.91). After reaching the peak, in each case, $u^*_z$ decreases to an equilibrium solution corresponding to the cross-flow strength. For the intermediate cross-flow cases, the systems first develop oblique 3D structures similar to the weak cross-flow cases, and then transition to longitudinal rolling vortices (for Couette flow at $u^*_{wall}$=2.24 and 2.40; for Poiseuille flow at $u^*_{center}$=2.84, 2.96, and 3.40), as shown in FIG. 8. For strong cross-flow (for Couette flow at $u^*_{wall}$=4; for Poiseuille flow at $u^*_{center}$=3.8) the flows develop directly into longitudinal 2D rolling patterns.

Unlike the evolution of the maximum $u^*_z$, $Ne$ always increases during the transition from 3D to 2D vortices (FIG. 7 c-d). However, when the cross-flow is not strong enough to suppress the 3D structures, the steady-state value of $Ne$ for the stronger cross-flow can be lower than in the weaker cross-flow. In the cross-flow with suppressed the transverse structures, the system yields a longitudinal rolling pattern with a constant $Ne$=1.41, independent of the strength or type of cross-flow. As with the $u_z$ analysis, the charge transport by the longitudinal vortices is simply superimposed onto the cross-flow regardless of the flow profile.

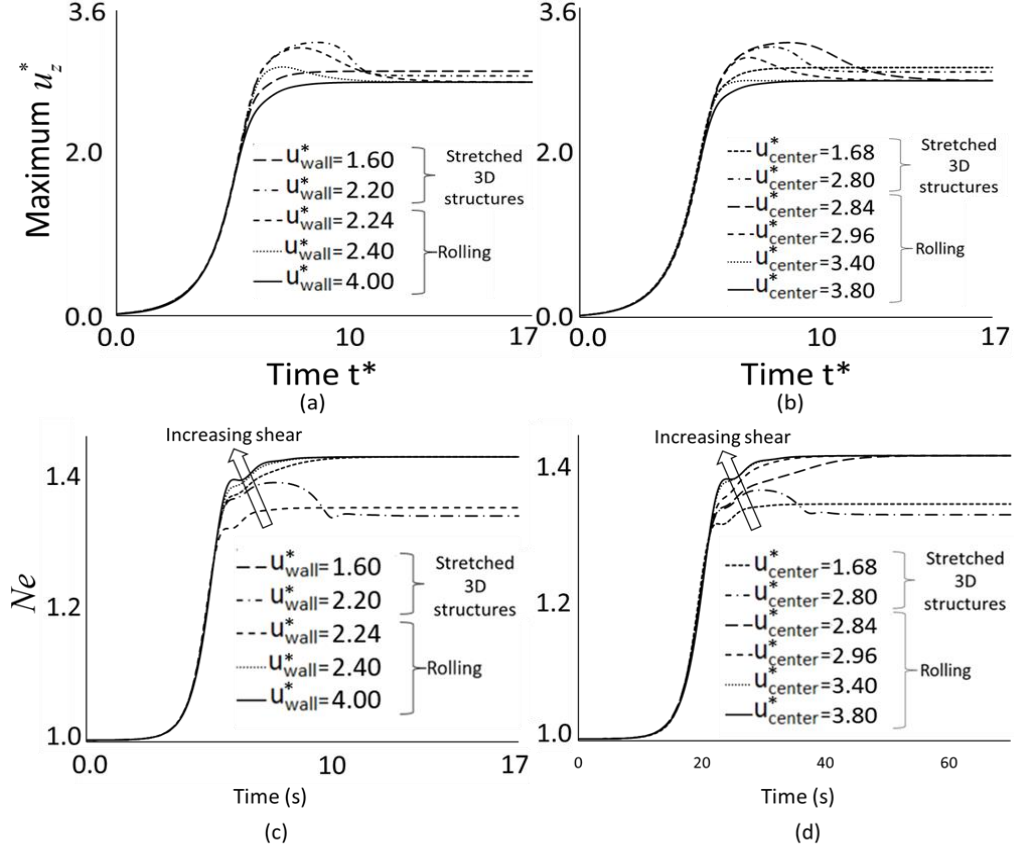

**FIG. 7.** Time evolution of maximum $u_z^*$ and $Ne$ for (a, c) Couette cross-flow and (b, d) Poiseuille cross-flow. Maximum $u_z^*$ have similar growth rates (~0.896) in the linear growth region ($t^*=0\sim5$). The square pattern initial perturbation scheme (Eq. (24)-(26)) is used. For strong cross-flow, the systems develop into longitudinal rolling patterns. For the weak cross-flow, the systems develop into oblique 3D structures with both transverse and longitudinal structures.

FIG. 8 shows $u_z^*$ at $z=H/2$ and $t^*=7.5$ for (a) Couette cross-flow with $u^*_{wall}=2.24$ and (b) Poiseuille cross-flow with $u^*_{center}=2.84$. At this time, the maximum $u_z^*$ reaches its peak value in the non-linear growth region. Both plots exhibit a dominating longitudinal rolling pattern aligned with the cross-flow in the x-direction. The transverse vortex is suppressed due to the interaction of the vortex's x-velocity components with the cross-flow; these interactions are most profound near the walls where x-velocity components of the initial 3D vortices are the greatest. For example, in Couette flow, the clockwise vortex of a vortex pair deforms at some oblique angle as the x-direction (streamwise) flow accelerates the upper region of the 3D structure and retards the bottom vortex region. This progress is reversed in the case of the counterclockwise rotating vortex of the pair. Eventually, these transverse structures become suppressed, and the systems develop into longitudinal rolling patterns [88]. Since the longitudinal rolling pattern is two dimensional in the y and z-directions, it does not interact with the bulk cross-flow. For the Poiseuille flow, the mechanism is slightly different; however, the interactions of the vortex structure and the bulk flow exist only in the x-direction; thus y-z structures are not affected by the cross-flow; therefore, the streamwise vortices cannot be suppressed by any type of the cross-flow.

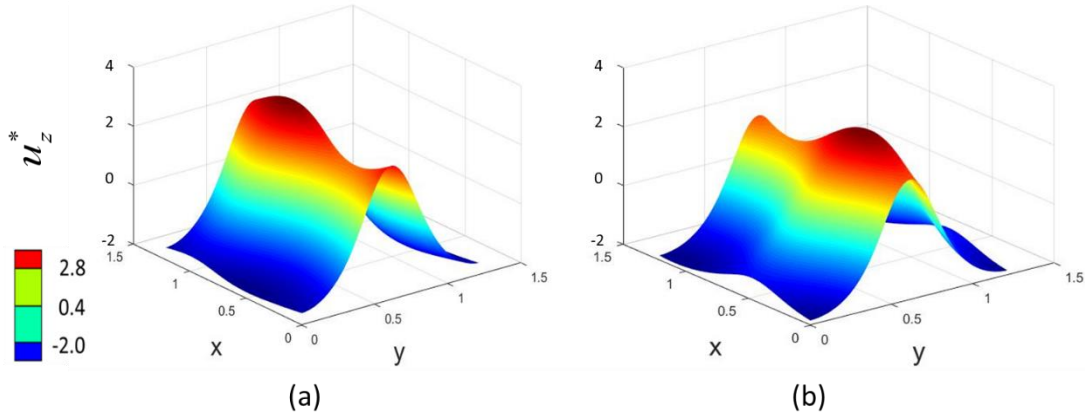

**FIG. 8.** Contours of $u_z^*$ for $z=H/2$, t=7.5: (a) Couette flow $u_{wall}^* = 2.24$; (b) Poiseuille flow $u_{center}^* = 2.84$. The color map corresponds to the values of $u_z$, which is also given by the vertical axis.

DMD analysis of the EC vortices in the cross-flow was performed using the numerical data of $u_z^*$ in the linear growth region (t*=0-7 for Couette flow and t*=0-6.25 for Poiseuille flow) at time intervals of $\Delta t^* = 0.125s$. A greater number of unstable dynamic modes exist in the oblique 3D pattern compared to the rolling pattern. FIG. 9 shows the $\lambda$ for Couette cross-flow at (a) $u^*_{wall}$=2.20 and (c) $u^*_{wall}$=2.24. Similarly, FIG. 10 shows $\lambda$ for Poiseuille cross-flow: (a) $u^*_{center}$=2.80 and (c) $u^*_{center}$=2.84. As in the $u_z^*$ evolution in hydrostatic base states without cross-flow (FIG. 4-FIG. 6), a perturbation in cross-flow arouses several unstable dynamic modes. Most of the dynamic modes are similar in the corresponding flows, resulting in a similar flow field up to the bifurcation point. However, in both cases, the lower velocity flow contains additional unstable modes, i.e., $m_1$ and the conjugate pair $m_2 - \bar{m}_2$ in Couette cross-flow, and mode $m_3 - \bar{m}_3$ in Poiseuille cross-flow. These additional unstable dynamic modes correspond to 3D features changing the stability of the system; they appear in the nonlinear growth region up to the bifurcation point where the curves of weak and strong cross-flow start to diverge (see FIG. 7(a-b)).

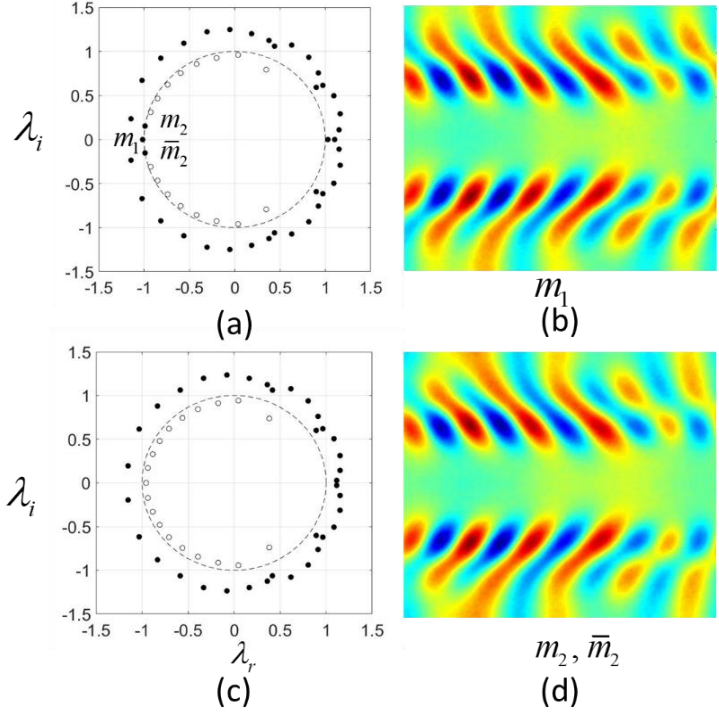

**FIG. 9.** Eigenvalues $\lambda_i$ for $u_z^*$ in linear growth region (t*=0-7) for Couette cross-flow ((a) $u^*_{wall}$=2.20 and (c) $u^*_{wall}$=2.24). Three additional unstable dynamic modes in $u^*_{wall}$=2.20 case change the equilibrium solution from a rolling pattern to oblique 3D structures. The corresponding eigenvectors sliced at $z = H/2$ are shown in (b) mode $m_1$ and (d) mode $m_2$ and $\bar{m}_2$.

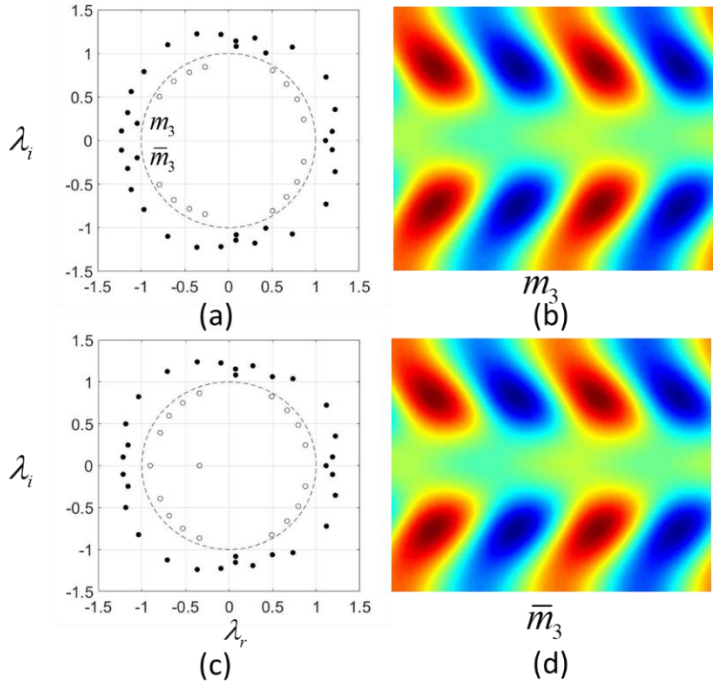

**FIG. 10.** Eigenvalues $\lambda_i$ for $u_z^*$ in linear growth region (t*=0-6.25s) for Poiseuille cross-flow ((a) $u^*_{center}$=2.80 and (c) $u^*_{center}$=2.84). An additional pair of conjugate unstable dynamic modes in $u^*_{center}$=2.80 case change the equilibrium solution from a rolling pattern to an oblique 3D structure pattern. The corresponding eigenvectors sliced at $z = H/2$ are shown in (b) mode $m_3$ and (d) mode $\bar{m}_3$.

## 4. Pattern transition after cross-flow application

This section studies the transitions of 3D to 2D patterns by applying the cross-flow to already developed square vortex structures, as shown in FIG. 3(a). With weak cross-flow, the systems transitions to oblique 3D vortex structures (oblique transverse and regular longitudinal structures coexist). The increased cross-flow yields a longitudinal rolling pattern, i.e., transverse structures are fully suppressed. FIG. 11 shows the time evolution of maximum $u_z^*$ in Couette cross-flow. For lower cross-flow velocities (e.g., $u^*_{wall}$=2.40) and, therefore, weak shear stress, the maximum $u_z^*$ decreases to an equilibrium value, that is somewhat greater than that for the rolling pattern flow ($u^*_{wall}$ =2.75). Interestingly, with further increase in $u^*_{wall}$ (e.g., $u^*_{wall}$=3.20), the equilibrium value of $u_z^*$ may decrease below the value of the rolling pattern. And with even further increasing $u^*_{wall}$ (e.g., $u^*_{wall}$=3.84), the equilibrium solution develops an oblique 3D structure with maximum $u_z$ that is greater than that of the rolling pattern. However, at $u^*_{wall}$ = 3.88, a bifurcation occurs, and the steady-state solution has only 2D streamwise vortices. The transition from 3D to the 2D rolling pattern is marked by a significant increase in $u_z^*$ to a value greater than the original square pattern, before finally decaying to the value of the rolling pattern. This significant increase is a result of kinetic energy transfer from modes with 3D structures to the dominating 2D structures. For larger $u^*_{wall}$, the peak $u_z^*$ value is reduced, and the time required for pattern transition also decreases. When the applied $u^*_{wall}$ is sufficiently large (e.g., $u^*_{wall}$>6.8), the maximum value of $u_z^*$ never is never above that of the rolling pattern.

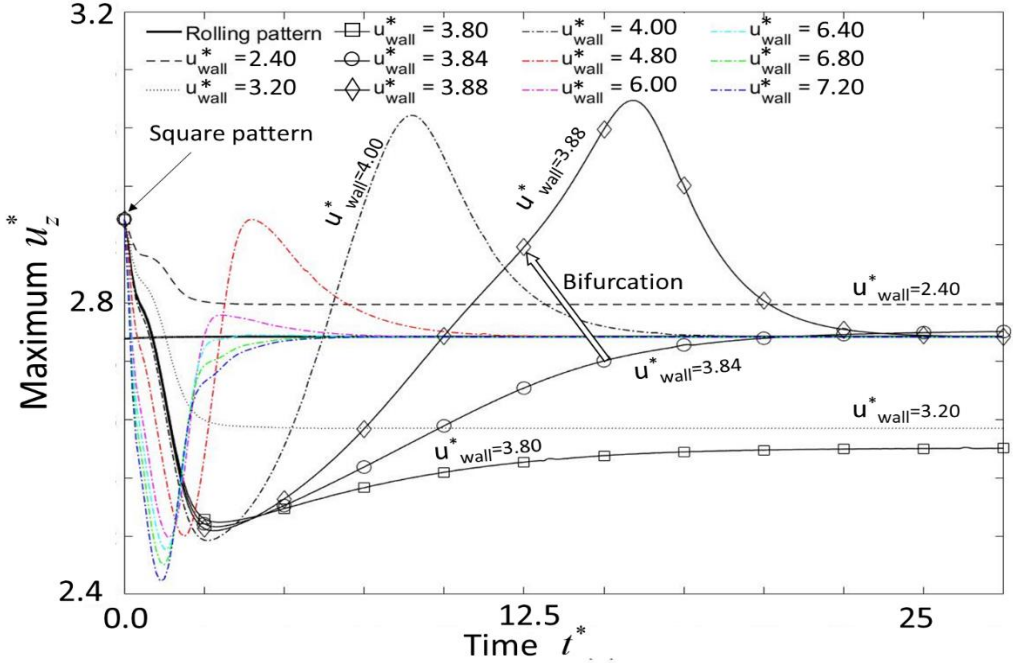

**FIG. 11.** Time evolution of maximum $u_z^*$ after a finite velocity is applied to the upper wall. For small $u^*_{wall}$, the maximum $u_z^*$ decreases and reaches a new equilibrium state where oblique 3D structures are observed. For large $u^*_{wall}$, the maximum $u_z^*$ decreases down to the rolling pattern where longitudinal rolls dominate, after a nonlinear transition. Bifurcation happens at $u^*_{wall}$=3.88.

Similar behavior is observed for the Poiseuille flow. FIG. 12 shows the time evolution of maximum $u_z$ due to an applied uniform body force $F_p$ used to obtain the Poiseuille flow. For small $u^*_{center}$ and, therefore, weak applied shear stress (e.g., $u^*_{center}$=3.40), the maximum $u^*_z$ decays to an equilibrium state with a value greater than that of the rolling pattern (2.75). With increasing $F_p$ (e.g., $u^*_{center}$=3.92), the equilibrium solution develops an oblique 3D structure as the maximum $u_z$ slightly increases after decaying to a minimum value. With $u^*_{center}$ up to 3.92, both oblique transverse and longitudinal structures coexist in the equilibrium solution. However, at $u^*_{center}$ =3.96, a bifurcation occurs, and the steady-state solution yields 2D streamwise vortices only. The transition from 3D to the 2D rolling pattern is marked by a significant increase in $u^*_z$ to a value greater than the original square pattern before ultimately decaying into the rolling pattern. For large $F_p$, the peak value of $u^*_z$ is reduced, and the time required for pattern transition also decreases. When the applied $F_p$ is sufficiently large (e.g., $u^*_{center}$=5.52), the maximum value of $u^*_z$ does not exceed the levels above that of the rolling pattern.

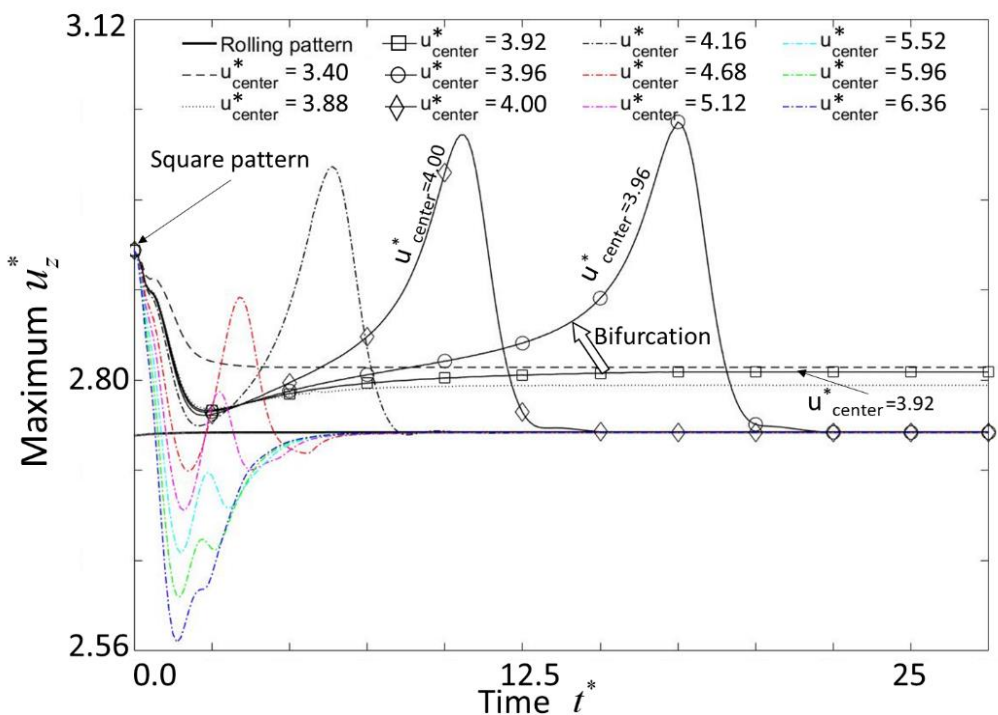

**FIG. 12.** Time evolution of maximum $u^*_z$ after a uniform body force is applied to the flow field. For small $u^*_{center}$, the maximum $u^*_z$ decreases and reaches a new equilibrium state where oblique 3D structures are observed. For large $u^*_{center}$, the maximum $u^*_z$ decreases to the rolling pattern values where longitudinal rolls dominate, after a nonlinear transition. The bifurcation occurs at $u^*_{center}$=3.96.

To analyze the coherent structures leading to the suppression of the instabilities, DMD analysis of the EC in the cross-flow was performed using the numerical results in the transition region for $\Delta t^* = 0.25$. FIG. 13 shows $\lambda_i$ for the weak cross-flow cases (a) $u^*_{wall}$=3.84 for Couette cross-flow, and (c) $u^*_{center}$=3.88 for Poiseuille cross-flow. The corresponding unstable eigenvectors correspond to the non-decaying coherent flow structures. In addition to the dominant dynamic modes ($m_4$ and $m_{10}$) corresponding to the rolling pattern, unstable dynamic modes ($m_5$-$m_9$ and $m_{11}$-$m_{12}$) exist; these are associated with the oblique 3D features. The unstable modes are similar to the ones obtained from the linear growth of perturbation in the cross-flow scenario, see FIG. 9 and FIG. 10, which can lead to the changes in the stability of the entire system. FIG. 14 shows the analysis of the strong cross-flow cases ($u^*_{wall}$=3.88 for Couette and $u^*_{center}$=4.16 for Poiseuille cross-flows). Only a single unstable eigenvalue is observed, which corresponds to rolling pattern eigenvectors $m_4$ or $m_{10}$ in FIG. 13. The DMD analysis is consistent with the numerical simulation; the EC flow transforms from 3D square to 2D rolling with the strong cross-flow. A list of the unstable eigenvalues is included in Table II of supplementary material.

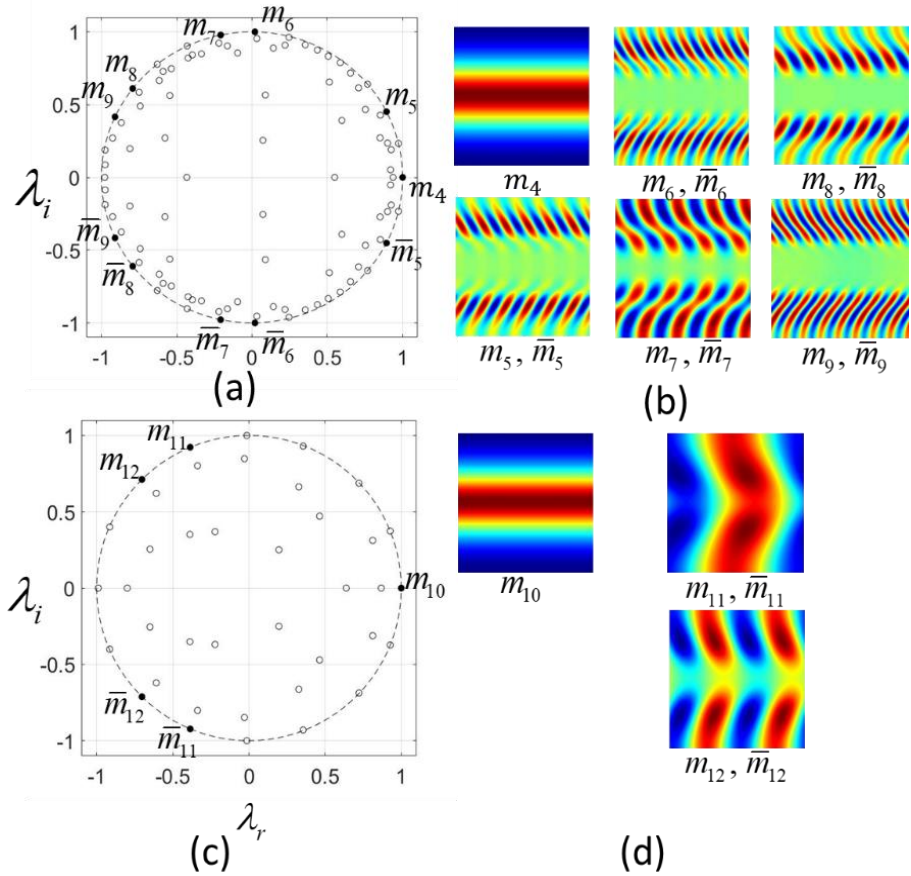

**FIG. 13.** Eigenvalues $\lambda_i$ for $u^*_z$ in transition region for Couette type cross-flow (a) $u^*_{wall}$=3.84 and Poiseuille type cross-flow (c) $u^*_{center}$=3.88. Unstable dynamic modes change the equilibrium solution from a square pattern to oblique 3D structures. The corresponding eigenvectors sliced at $z = H/2$ are shown in (b) mode $m_4 - \overline{m}_9$ and (d) mode $m_{10} - \overline{m}_{12}$.

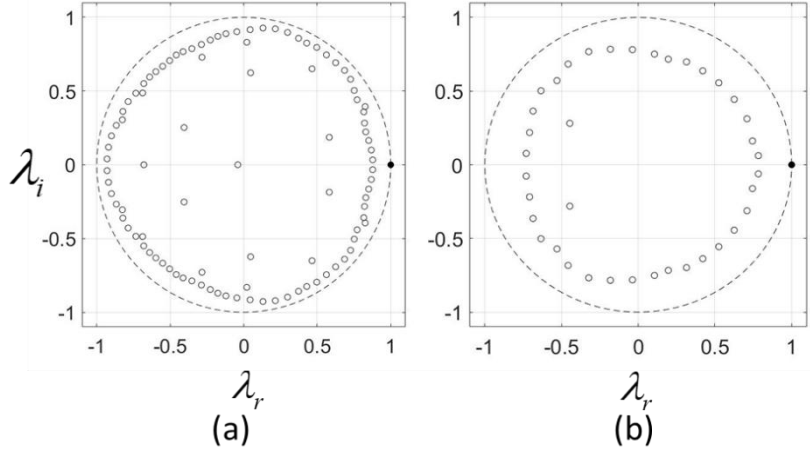

**FIG. 14.** Eigenvalues $\lambda_i$ for $u_z^*$ in transition region for Couette type cross-flow (a) $u^*_{wall}$=3.88 and Poiseuille type cross-flow (c) $u^*_{center}$=4.16. The unstable dynamic modes yield the rolling structures.

FIG. 15 shows $u_z^*$ at $z = H/2$ for Couette and Poiseuille cross-flow when the maximum value reaches the valley (t$^*$=2.5, $u^*_{wall}$=3.88 and t$^*$=2, $u^*_{center}$=4.16) and peak (t$^*$=15.75, $u^*_{wall}$=3.88 and t$^*$=6.25, $u^*_{center}$=4.16), as shown in FIG. 11 and FIG. 12. For both types of cross-flow, the $u_z^*$ patterns are similar. When the maximum $u_z^*$ is at its valley, oblique 3D structures are more pronounced, while a rolling pattern dominates the flow for high maximum $u_z^*$. The transition can be interpreted as energy transfer from one dominant mode to another. Further analysis of the nonlinear transition behavior can be performed by solving for a reduced nonlinear system such as given by the coupled Ginzburg-Landau equations for transverse and longitudinal rolls, similar to the analysis of the effects of cross-flow on RBC [66-68].

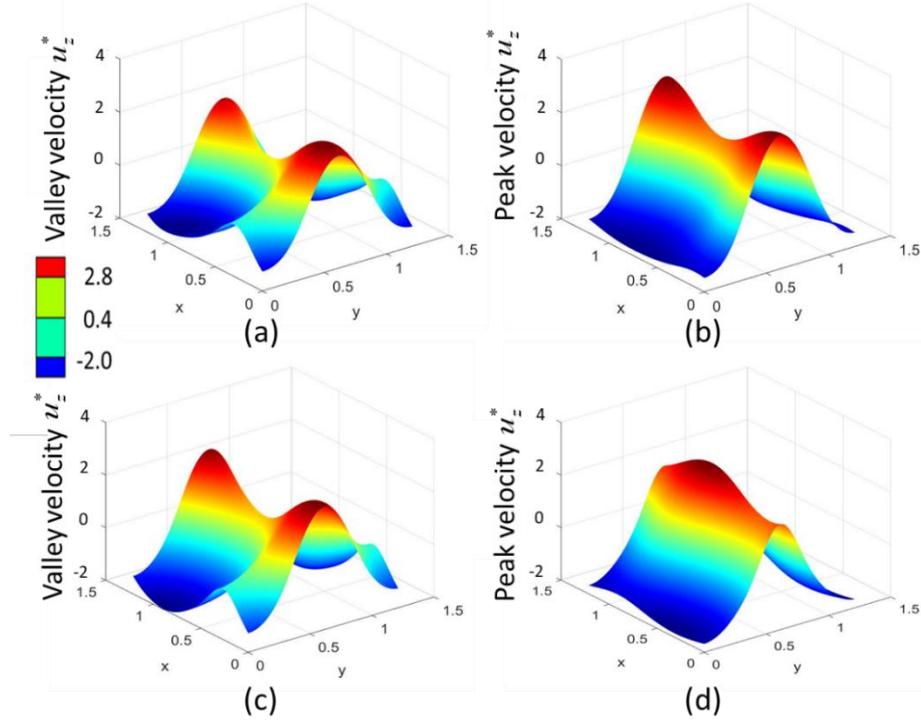

**FIG. 15.** Contours of valley and peak velocity $u_z^*$ for (a-b) Couette cross-flow ($u^*_{wall}$=3.88) and (c-d) Poiseuille cross-flow ($u^*_{center}$=4.16)

FIG. 16 shows the iso-surfaces of charge density during the transition from a square to a rolling pattern. Square patterns of charge density are observed at the conditions without cross-flow, as shown in FIG. 16(a). When a weak cross-flow is applied, the iso-surfaces are obliquely stretched in the x-direction, as shown in FIG. 16(b-c). For strong cross-flow, the transverse patterns are suppressed, and only a rolling pattern is observed, see FIG. 16(d). The iso-surface of charge density is identical for all strong cross-flow cases, and the rolling pattern perturbation without cross-flow.

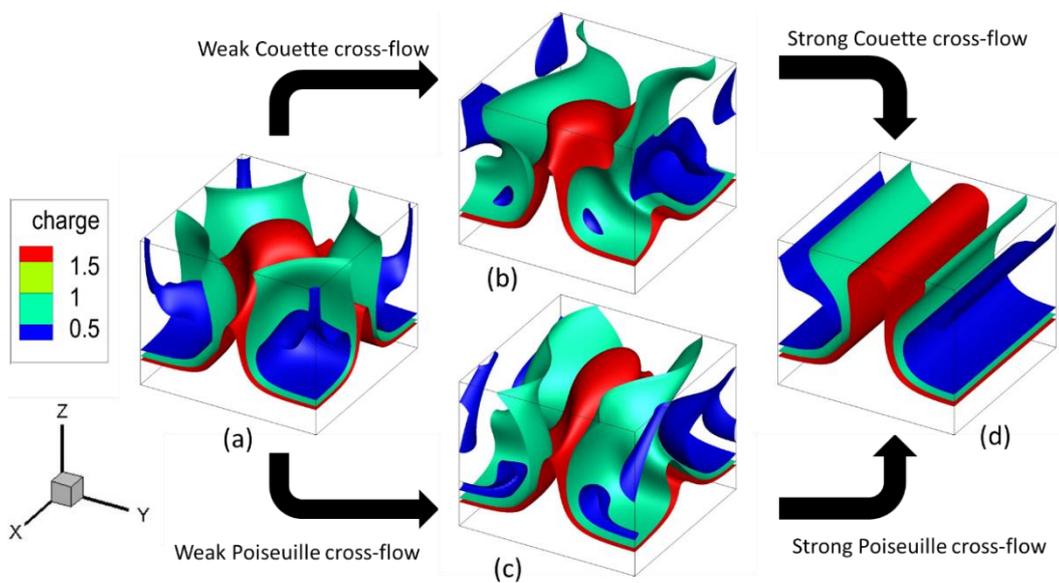

**FIG. 16.** Iso-surfaces of charge density for (a) square pattern without cross-flow, (b) $u^*_{wall}$=3.84, (c) $u^*_{center}$=3.92 and (d) strong cross-flow/rolling pattern.

FIG. 17 and FIG. 18 show the dependence of the electrical Nusselt number on the non-dimensional parameter $Y$ calculated in the x-direction. Hysteresis behavior with well-defined bifurcation is observed for both Couette and Poiseuille cross-flows. The bifurcation thresholds are $Y_c = 772.73$, $Y_f = 438.14$ for Couette cross-flow and $Y_c = 300.75$, $Y_f = 213.90$ for Poiseuille cross-flow. At $Ne = 1$ (base state) EC vortices are not present. If $Y > Y_c$, the square pattern perturbation (Eq. (24)-(26)) results in oblique 3D structures. For $Y < Y_c$, any perturbation results in streamwise rolling vortices as the equilibrium solution. Oblique 3D flow features develop when shear stress is applied to the square pattern. As $Y$ is reduced (shear stress increased), the oblique features persist until $Y=Y_f$; additional reduction in $Y$ suppresses the features in the transverse direction -- only the longitudinal structures are possible. The $Ne$ value is lower in 3D EC vortices; the oblique 3D structures result in decreasing $Ne$. When $Y$ is close to $Y_f$, $Ne$ slightly increases before transitioning to the rolling pattern value. This increase of $Ne$ agrees with trends in maximum $u_z$, as shown in FIG. 11 and FIG. 12. The inserts in FIG. 17 and FIG. 18 show $u_z$ contour plots at $z = H/2$; the $u_z$ profiles are oblique in the cross-flow direction for both Couette and Poiseuille cross-flows. The hysteresis loop can be closed by introducing a y-directional cross-flow to suppress the rolling pattern vortices [88].

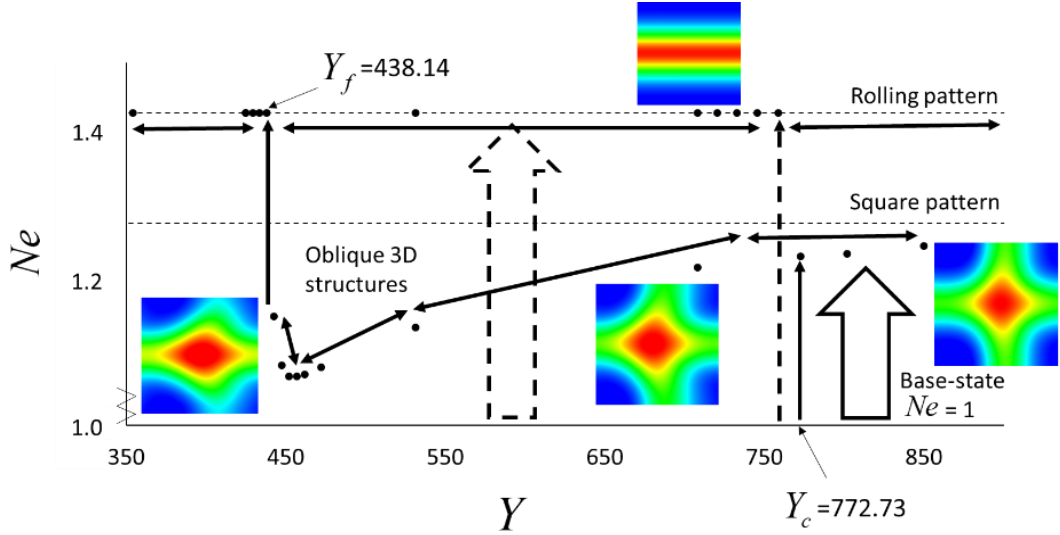

**FIG. 17.** Hysteresis loop of $Ne$ vs. $Y$ for Couette cross-flow. The bifurcation thresholds are $Y_c = 772.73$, $Y_f = 438.14$.

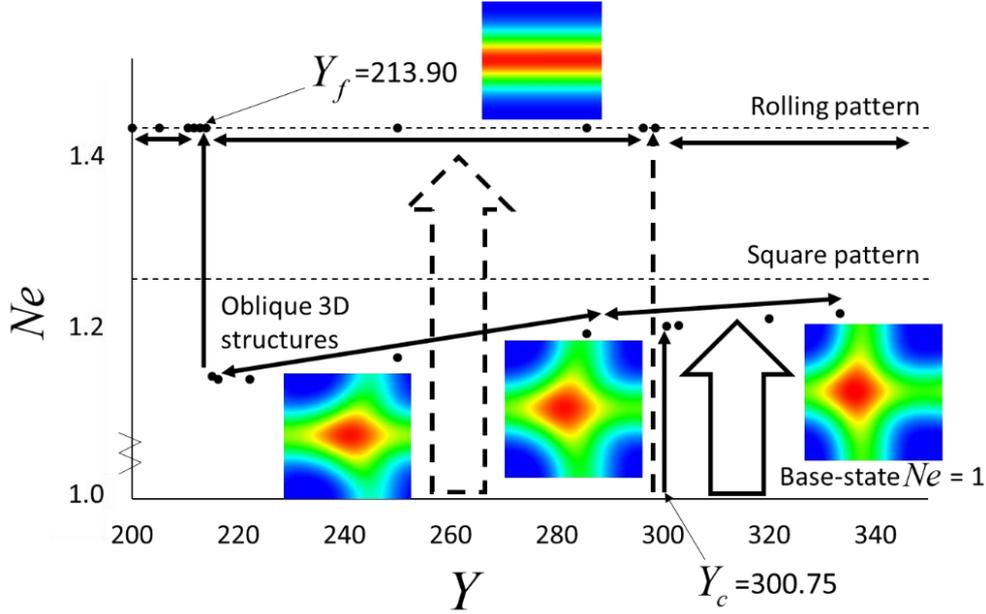

**FIG. 18.** Hysteresis loop of $Ne$ vs. $Y$ for Poiseuille type cross-flow. The bifurcation thresholds are $Y_c = 300.75$, $Y_f = 213.90$.

## VI. CONCLUSION

The 3D numerical study extends the EC stability analysis of 3D flow structures in Couette and Poiseuille cross-flows between two infinite parallel plates in the EHD flow with unipolar charge injection. The numerical modeling approach uses a second-order TRT-LBM scheme to solve the flow and charge transport equations coupled to a Fast Poisson Solver for the electrical potential. Shear containing cross-flow first stretches the EC cells at obliques angles due to the interaction between the vortices and the bulk flow. These interactions form oblique 3D features before transitioning to 2D streamwise vortices after a sufficiently high cross-flow velocity is reached. The transition from 3D to 2D equilibrium states is observed for all initial perturbation schemes and independent of the domain configurations considered in this work, i.e., square perturbations and it's harmonic, oval, hexagonal, and mixed perturbations. Two transitional scenarios are studied, (i) the cross-flow is applied before and (ii) after the EC vortices are established. If the cross-flow is applied before the perturbation leading to the formation of EC vortices, bifurcation occurs at $u^*_{max}$=2.20 for Couette flow and $u^*_{center}$=2.80 for Poiseuille flow. If the cross-flow is applied after the EC vortices are established, the convective cells are more stable; the bifurcation occurs at $u^*_{max}$=3.88 for Couette flow and $u^*_{center}$=3.96 for Poiseuille flow.

DMD analysis is applied in both the linear growth and the nonlinear transition regions, providing insight into the development of the coherent flow structures, predicting the linear behavior, and identifying bifurcation thresholds. The dynamic modes obtained from the linear growth region agree with the global growth rates obtained from the evolution of $u^*_z$. The DMD based predictions agree with the simulation results and can be used to accelerate the computational process in the linear growth region of systems such as EC, RBC, and magneto-convection. The bifurcation thresholds between oblique 3D structures and the rolling pattern are characterized by the presence of additional unstable dynamic modes in the weak cross-flow cases responsible for the development of oblique 3D flow features.

To parameterize the effect of cross-flow on EC vortices, the non-dimensional analysis of the governing equations uses a parameter $Y$ which arises in the momentum equation in the streamwise direction. Similar to the 2D cases where cross-flow suppresses convective vortices [88], a hysteresis in 3D cases is observed. The bifurcation thresholds are $Y_c = 772.73$, $Y_f = 438.14$ for Couette flow and $Y_c = 300.75$, $Y_f = 213.90$ for Poiseuille flow. The applied shear organizes the flow into 2D rolls parallel to the mean flow and enhances the convection marked by a significant increase in *Ne*, similar to heat convection problem with moderate Rayleigh number [102]. The presented approach can be applied to other convection flow systems such as electro-kinetic instabilities in overlimiting current region, Rayleigh-Benard convection, Marangoni effects, and magneto-convection with extended non-linear perturbation schemes such as triangular or hexagon patterns.

## VII. ACKNOWLEDGMENTS

This research was supported by the DHS Science and UK Home Office; grant no. HSHQDC-15-531 C-B0033, by the National Institutes of Health, grant NIBIB U01 EB021923 and NIBIB R42ES026532 subcontract to UW.